\documentclass[11pt,a4paper]{article}
\usepackage{jheppub}

\usepackage{epsfig}
\usepackage{amsmath}
\usepackage{amssymb}
\usepackage{amsbsy}
\usepackage{enumerate}
\usepackage{bbm}
\usepackage{url}
\usepackage[hang,sf]{subfigure}
\def\beq{\begin{equation}}
\def\beqn{\begin{eqnarray}}
\def\eeq{\end{equation}}
\def\eeqn{\end{eqnarray}}

\def\({\left(} 
\def\){\right)}

\newdimen\figwidth
\figwidth=\textwidth

\newcommand\POWHEG{{\tt POWHEG}}
\newcommand\HERWIG{{\tt HERWIG}}
\newcommand\HWpp{{\tt Herwig++}}
\newcommand\PYTHIA{{\tt PYTHIA}}

\newcommand\BOX{{\tt POWHEG BOX}}
\newcommand\MCatNLO{{\tt MC@NLO}}
\newcommand\aMCatNLO{{\tt aMC@NLO}}
\newcommand\madfks{{\tt MadFKS}}
\newcommand\MadGraph{{\tt MadGraph4}}

\newcommand\sss{\mathchoice%
{\displaystyle}%
{\scriptstyle}%
{\scriptscriptstyle}%
{\scriptscriptstyle}%
}

\newdimen\hbigcirc
\newdimen\wbigcirc

\newcommand\pt{p_{\sss\rm T}}

\newcommand\kt{k_{\sss\rm T}}

\newcount\minutes 
\newcount\scratch 
 
\def\timestamp{
\scratch=\time 
\divide\scratch by 60 
\edef\hours{\the\scratch} 
\multiply\scratch by 60 
\minutes=\time 
\advance\minutes by -\scratch 

\today\ --$\,$\hours:\null 
\ifnum\minutes< 10 0\fi 
\the\minutes
} 

\def\nn{\nonumber}

\title{Single-top $\boldsymbol{t}$-channel hadroproduction in the
  four-flavour scheme with {\tt\bf POWHEG} and {\tt\bf aMC@NLO}}
\author[a]{Rikkert Frederix,}
\author[b]{Emanuele Re,}
\author[c,d]{Paolo Torrielli.}
\affiliation[a]{Institute for Theoretical Physics, University of Z\"urich,\\
  Winterthurerstrasse 190, CH-8057 Z\"urich, Switzerland}
\affiliation[b]{Institute for Particle Physics Phenomenology, Department of Physics\\
  University of Durham, Durham, DH1 3LE, UK}
\affiliation[c]{ITPP, EPFL,\\
  CH-1015 Lausanne, Switzerland}
\affiliation[d]{PH Department, TH Unit,\\
  CERN, CH-1211 Geneva 23, Switzerland}
\emailAdd{frederix@physik.uzh.ch}
\emailAdd{emanuele.re@durham.ac.uk}
\emailAdd{paolo.torrielli@epfl.ch}
\abstract{ We present results for the QCD next-to-leading order (NLO)
  calculation of single-top $t$-channel production in the 4-flavour
  scheme, interfaced to Parton Shower (PS) Monte Carlo programs
  according to the \POWHEG{} and \MCatNLO{} methods. Comparisons
  between the two methods, as well as with the corresponding process
  in the 5-flavour scheme are presented. For the first time results
  for typical kinematic distributions of the spectator-$b$ jet are
  presented in an NLO+PS approach.}
\begin{document}
\maketitle

\section{Introduction}
\label{sec:introduction}
Although experimentally challenging, the electroweak production of top
(or antitop) quarks without their antiparticles, known as single-top
production, is particularly important, not only because it provides a
relatively clean place to study the electroweak properties of the top
quark~\cite{Bernreuther:2008ju}, but also because it allows for a
direct measurement of
$V_{tb}$~\cite{Alwall:2006bx,Lacker:2012ek}. Moreover, single-top
production is sensitive to many BSM
models~\cite{Tait:2000sh,Cao:2007ea}, that in some cases are difficult
to discover in other search channels. Single-top production is in
general also a background for all searches where top-pair production
and the production of a $W$ boson in association with jets are an
important background.

At leading order it is possible to classify the single-top
hadroproduction mechanisms according to the virtuality of the
$W$ boson appearing in the tree-level diagrams: these
mechanisms are known as $s$-, $t$-, and $Wt$-channel processes.  The distinction becomes ill-defined at higher orders because of
interference effects.\footnote{A notable exception is represented by
  the possibility of treating $s$ and $t$ channels separately when
  including QCD NLO corrections in the 5-flavour scheme (that will be
  shortly described later). The exact definition of the $Wt$ channel
  is instead more complicated.} However, it makes sense to keep this
nomenclature, since in general the kinematic properties of the final-state objects are very different when one includes higher-order
corrections to $s$-, $t$- or $Wt$-channel
Born-level processes.  Moreover, preserving this distinction is also
important because different BSM scenarios affect the three production mechanisms differently, making single-top studies a promising approach to
distinguish among New-Physics models.  Not surprisingly,
experimental results and prospects for new analysis strategies at the
Tevatron and LHC, as well as the aforementioned theoretical
computations, are very often presented keeping this distinction. In
this paper we will only concentrate on the main production mechanism, i.e.~the $t$ channel.

The theoretical effort to produce accurate predictions for this
process is significant, and several improvements with respect to the
first result of ref.~\cite{Willenbrock:1986cr} were achieved in the
past: after the QCD NLO predictions obtained in
refs.~\cite{Harris:2002md,Campbell:2004ch,Cao:2005pq}, studies on the
impact of resummation~\cite{Wang:2010ue,Kidonakis:2011wy}, off-shell
effects~\cite{Falgari:2010sf,Falgari:2011qa}, and electroweak
corrections~\cite{Beccaria:2008av} were also performed in recent
years. All these results were based on the 5-flavour-scheme
description of the process. In 2009 the first results for the NLO
predictions in the 4-flavour scheme became
available~\cite{Campbell:2009ss,Campbell:2009gj}. More recently,
factorizable NLO corrections to the top-quark decay process were
computed and included in the \texttt{MCFM} framework~\cite{Campbell:2012uf},
allowing for a consistent treatment of spin-correlation effects
between the production and the decay stage, in the zero-width
approximation for the top quark.

Single-top production is difficult to be observed as a signal itself,
essentially because of the huge $W$+jets and the large $t\bar{t}$
backgrounds, which are hard to reduce by means of standard techniques,
such as imposing cuts. As a consequence, to discriminate between
signal and backgrounds and claim the evidence and the observation of
single-top production, as well as to measure the $s$- and $t$-channel
production rates, techniques based on neural networks, boosted
decision trees and multi-variate analysis techniques were used at the
Tevatron~\cite{Aaltonen:2010jr,Abazov:2009ii,Abazov:2009pa,Group:2009qk},
and are currently employed at the LHC as well, on top of the cut-based
ones~\cite{Chatrchyan:2011vp,Aad:2012gd,Aad:2012ux,Aad:2012dj}.  Since
all these strategies rely on the Monte-Carlo modelling of
the signal and backgrounds, it is clear that the Monte-Carlo tools
play a crucial role in understanding single-top production at hadron
colliders.  Therefore it is highly desirable to have event generators
that consistently incorporate as much as possible of the improvements
achieved in the theoretical predictions. An obvious but nontrivial
direction to pursue is that of including QCD NLO corrections into Parton
Shower (PS) Monte Carlo event generators. This has been achieved with
the \MCatNLO{} and the \POWHEG{}
methods~\cite{Frixione:2002ik,Nason:2004rx}, so that nowadays it is
possible to simulate with NLO+PS accuracy several processes relevant for lepton
and hadron colliders, as well as deep-inelastic scattering.  The
recent automation of these
techniques~\cite{Alioli:2010xd,Frederix:2011zi,Hoche:2010pf,Hoeche:2011fd,Platzer:2011bc}
allowed the simulation of processes with many external massless and
massive partons and with non-trivial colour
structures.

The simulation of $s$-, $t$-, and $Wt$-channel single-top production
using the \MCatNLO{} and the \POWHEG{} methods is already possible, as
described in
refs.~\cite{Frixione:2005vw,Frixione:2008yi,Alioli:2009je,Re:2010bp}.
However, for the $t$-channel case, which was implemented following the
5-flavour scheme, there is some room for improvements.  In particular,
it is known that the modelling of the so called ``spectator $b$'',
i.e.~the hardest $b$ jet not originating from the top-quark decay, is
not very solid.  The kinematic properties of this $b$-flavoured jet
are, however, important to study single-top production in the $t$ channel, since
one usually requires one $b$ jet with high
and one with low transverse momentum (or, similarly, an anti-$b$ tag)
to tag a candidate single-top event as opposite to a $t\bar{t}$ pair,
where typically two $b$ jets with large momentum are present.  For
similar reasons, the kinematics of $b$-flavoured objects plays an
important role in the discrimination between $s$- and $t$-channel
processes. It is therefore desirable, especially for the precision
that the LHC will achieve, to have tools as accurate as possible in
describing these features.

To obtain NLO accuracy for the spectator-$b$ observables in a NLO+PS
program, a description of the single-top $t$-channel production
process in the 4-flavour scheme~\cite{Campbell:2009ss,Campbell:2009gj}
is required. The aim of this paper is to present results for the first
implementation of this process in the \MCatNLO{} and \POWHEG{}
frameworks.

\begin{figure}[htb]
  \begin{center}
    \epsfig{file=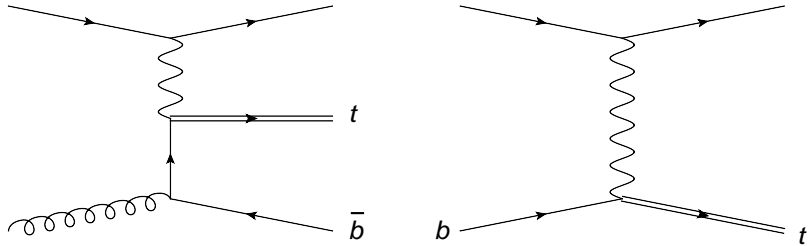,width=0.7\textwidth}\\
  \end{center}
  \caption{\label{fig:4f5ftch_lo} Representative leading-order Feynman
    diagrams for the single-top $t$-channel production in the 4-
    (left) and 5-flavour (right) scheme.}
\end{figure}

Let us elaborate on the differences and similarities between the 4-
and 5-flavour schemes, whose corresponding representative
leading-order Feynman diagrams are depicted in
fig.~\ref{fig:4f5ftch_lo}.

The difference is, in short, that in the former the PDF does not
contain $b$ quarks: all $b$ quarks are generated in the matrix
elements and cannot directly come from the (anti)proton. In the latter
scheme, the description of the PDF contains the $g\to b\bar{b}$
splitting.  This leads to the following considerations.

\begin{itemize}
\item In the 5-flavour scheme, the mass of the (initial-state) $b$
  quark has to be neglected in the matrix elements for factorisation
  to be valid. The cross section depends on the $b$
  mass ($m_b$) only through the starting scale for the evolution of the $b$
  quark in the PDF. On the other hand, in the 4-flavour scheme, the
  mass appears explicitly in the matrix elements, while the PDFs are
  independent of the $b$ quark. In ref.~\cite{Campbell:2009gj} it has
  been shown that the $b$-mass dependence of the total cross section
  is similar in size in both schemes.
\item In the 5-flavour scheme the PDF evolution resums logarithms of
  the form $\log(\mu_F^2/m_b^2)$, where $\mu_F$ is the factorisation
  scale, and the radiation thus resummed is described at
  leading-logarithmic accuracy. The clear advantage is that due to the
  resummation the total rate can be predicted precisely.  In the
  4-flavour scheme, these logarithms are not resummed. However,
  the non-logarithmic contributions to the $g \to b \bar{b}$ splitting
  are already taken into account in the LO description of the process,
  which leads to a more precise prediction of the $b$-quark
  distributions than in the 5-flavour scheme. When including NLO
  corrections in the 4-flavour scheme, the first logarithms are
  included as well. Similarly, at the NLO level in the
  5-flavour scheme, also the first non-logarithmic contributions due
  to radiation are correctly accounted for.
\item The running of the strong coupling $\alpha_s$ depends on the
  number of flavours.
\end{itemize}

The $m_b=0$ approximation in the 5-flavour scheme can be improved
systematically by replacing (higher-order) contributions that have no
$b$ quarks in the initial state, by corresponding contributions in
which $m_b\ne 0$. With this improvement the 4- and 5-flavour-scheme
descriptions become exactly equivalent when all orders in perturbation
theory are included.

In refs.~\cite{Campbell:2009ss,Campbell:2009gj} it was shown that with
a proper physically-motivated choice of factorisation and
renormalisation scales, the agreement of the two approaches is good
already at the NLO approximation. Therefore, the 4-flavour-scheme
calculation, which has the obvious advantage of a more precise
treatment of the spectator-$b$ quark, is the preferred one
for an exclusive description of the process. This statement is confirmed by the conclusions of ref.~\cite{Maltoni:2012pa}, where the issue of choosing a $4$- or a $5$-flavour scheme for key processes
at hadron colliders, including single-top, is addressed. There it is indeed shown that the size of the initial-state logarithms, $\log(Q^2/m_b^2)$, that are resummed in the $b$ PDF within
the 5-flavour scheme is modest, except at large Bjorken $x$'s. Moreover, the effective scale $Q^2$ that enters these
logarithms is typically smaller than the hard scale of the process, being
accompanied by universal phase-space suppression factors, which further reduces the size of these logarithms.

This paper is organised as follows. In sec.~\ref{sec:description} we
briefly describe some technicalities of the computer programs used to obtain
the results presented in the following. In sec.~\ref{sec:results} we
show typical distributions relevant for single-top studies, comparing
the pure-NLO results with those obtained with our NLO+PS codes.  We
compare different prescriptions for the matching
(\POWHEG{}~vs~\MCatNLO{}) as well as different showering models
(\PYTHIA{}~vs~\HERWIG{}). Scale dependence and PDF errors are displayed for the results obtained with \aMCatNLO{}, while a full study encompassing all other sources of uncertainty, \textit{e.g.}~the values used for the $b$- and
  the top-quark masses, is beyond the scope of this work and has not
  been attempted. Instead, we perform a comparison with the
results obtained in the 5-flavour scheme, in particular for the
spectator-$b$ jet.  Finally, in sec.~\ref{sec:conclusions} we give our
conclusions.

\section{Description of the implementation}
\label{sec:description}
To simulate the 4-flavour single-top $t$-channel
production with NLO+PS accuracy, the \BOX{} package~\cite{Alioli:2010xd} and the \aMCatNLO{} framework~\cite{Frederix:2011zi} have been
used. In this section we shortly discuss the technical aspects of the
codes employed to obtain the results presented in this
work.

The \aMCatNLO\ code is a tool to compute processes at NLO accuracy
matched to parton showers using the
\MCatNLO\ method~\cite{Frixione:2002ik}. It is completely automatic
and general, its only limitation being represented by CPU
availability. It is built upon the
\madfks~\cite{Frederix:2009yq} framework which uses the FKS
subtraction scheme~\cite{Frixione:1995ms} to factor out the poles in
the phase-space integration of the real-emission squared
amplitudes. In this work, we do not use \texttt{MadLoop}~\cite{Hirschi:2011pa}
to evaluate the virtual corrections, and exploit the analytic results
of ref.~\cite{Campbell:2009ss} instead.

The \BOX{} is a program that automates all the steps described in
ref.~\cite{Frixione:2007vw}, turning a NLO calculation into a
\POWHEG{} simulation. The details of how the program works have been
largely described in ref.~\cite{Alioli:2010xd}, and therefore will not
be repeated here. To implement the NLO computation in the
\BOX{} program, we obtained the needed inputs as follows.
\begin{itemize}
\item The Born kinematics was obtained mapping the $2\to 3$ phase space
  to a multidimensional hypercube where the random numbers are
  generated.  We have built the phase space using the
  Bycling-Kajantie~\cite{Byckling:1969sx} parameterisation.
\item The tree-level $~2\to 3~$ and $~2\to 4~$ matrix elements were
  obtained using \MadGraph~\cite{Alwall:2007st}.
\item The same virtual corrections as computed in
  ref.~\cite{Campbell:2009ss} (and implemented in the \texttt{MCFM} package)
  have been used here as well. We have also checked the numerical value for
  the Born and the 1-loop finite part against the value in
  Appendix 2.8 of ref.~\cite{Hirschi:2011pa}.
\item Although at the Born level there are 5 coloured partons, the
  colour-linked squared amplitudes for this process factorise the
  full Born squared amplitudes: no colour correlation is possible
  between the light and the heavy current, since a $W$ boson is exchanged.
  For the same reason, the assignment of planar colour connections
  to the $2\to 3$ processes is trivial.
\item The spin-correlated squared amplitudes were computed using
  \texttt{FeynCalc}~\cite{Mertig:1990an}, and checked by comparison with the
  collinear limits of the corresponding $~2\to 4~$ radiative
  corrections.
\end{itemize}
The complete NLO implementation has also been extensively checked with
the implementation available in \texttt{MCFM}, which makes use of the
Catani-Seymour subtraction scheme~\cite{Catani:1996vz}. Agreement was found for total rates
and for several distributions.

\section{Results}
\label{sec:results}
In this section we present our results at different levels of
sophistication, comparing the fixed-order partonic predictions with those after
the showering and the hadronisation stage.  We will show only results
for top production at the Tevatron and at the LHC (with hadronic
center-of-mass energy $\sqrt{S}=8$ TeV), although the
generation of anti-top events was also implemented and tested,
leading to similar results as those presented here. At the matrix-element
level the top quarks have been assumed to be stable: the decays have been generated by the Shower Monte Carlo programs and forced to be
semi-leptonic ($t\to b\, \ell^+\nu$). Branching ratios have been set
to 1, so that plots are normalised to the total cross section.

We have chosen renormalisation and factorisation scales following the
arguments presented in ref.~\cite{Campbell:2009ss}: the scale for this
process is not determined by the heavy top quark, but rather by the
(maximal) transverse momentum of the $b$ quark. We have therefore opted for
\begin{equation}
\label{eq:centralscale}
\mu_R=\mu_F= 4 \sqrt{(m_b^2+p_{T,b}^2)}
\end{equation}
as our central scale choice.\footnote{In
  refs.~\cite{Campbell:2009ss,Campbell:2009gj} the scale choices for
  the heavy- and light-quark lines were different ($\mu_h=m_t/4$ and
  $\mu_l=m_t/2$, respectively) and it was found that the total scale
  dependence is dominated by that coming from the heavy-quark line. We
  have therefore opted for a scale based on the kinematics of the
  heavy-quark line, which reflects the properties of the spectator-$b$
  quark better than a fixed scale. We expect that the results
  presented here give a better description with respect to
  refs.~\cite{Campbell:2009ss,Campbell:2009gj} when the spectator-$b$
  quark has a large transverse momentum, and that no other observables
  are hampered, because $4 \sqrt{(m_b^2+p_{T,b}^2)}\sim m_t/4$ for the
  bulk of the events.}  We have also checked that the choice in
(\ref{eq:centralscale}) is numerically close to the scale $\mathcal
Q(z)$ proposed in ref.~\cite{Maltoni:2012pa} in the kinematic range
relevant for the results presented in the following. For the \POWHEG{}
results, the $b$-quark momentum is understood to be taken at the
underlying-Born level. We have used the {\tt MSTW2008(n)lo68cl\_nf4}
sets~\cite{Martin:2009iq} of parton-distribution functions for the
(N)LO predictions (and the corresponding sets for the 5-flavour
results presented in sec.~\ref{sec:comparsion_4_5}), with default
values and running of $\alpha_s$ as given by the
LHAPDF~\cite{Whalley:2005nh} interface.

The values of physical parameters entering the computation are as
follows:
\begin{eqnarray}
 &~& m_t=172.5\ \mbox{GeV},\,\ \ \ \ \ \ \ \ m_b=4.75\ \mbox{GeV},\,\ \ \ \ \ \ \ \ m_W=80.398\ \mbox{GeV},\,\nn\\
 &~& \alpha^{-1}=132.338,\, \ \ \ \ \ \ \ \ \ \ \ \sin^2\theta_W=0.22265,\,\nn
\end{eqnarray}
and a diagonal CKM matrix has been employed.

Jets have been defined according to the inclusive $\kt$
algorithm~\cite{Catani:1993hr}, as implemented in the {\tt FastJet}
package~\cite{Cacciari:2005hq,Cacciari:2011ma}, setting $R=0.7$ and
imposing a $20\, (30)$~GeV lower cut on jet transverse momenta for the
Tevatron (LHC). Moreover, jets are required to have pseudorapidity 
$|\eta_j|<2.5$. The jets containing at least one $b$-flavoured
particle (a $b$ quark at the parton level, one or more $B$ hadrons for
showered results) are called ``$b$-flavoured jets'', as opposite to the
others, denoted ``light jets''.

We have run \HERWIG{}~6.520~\cite{Corcella:2000bw} and
\PYTHIA{}~6.4.22~\cite{Sjostrand:2006za} with default parameters
(i.e. without using a particular tune), and with
multiple-parton interactions and underlying event switched
off. \PYTHIA{} has been run with the $\pt$-ordered shower. To simplify
the analysis, we have set the lighter $B$ hadrons, the muons, and the
pions stable.

In order to estimate PDF and scale dependence for the \aMCatNLO\ results, we
have used the reweighting method described in
ref.~\cite{Frederix:2011ss}, which allows the extraction of these
theoretical uncertainties without any extra CPU costs.  For the PDF
dependence the events are reweighted by the $2 \times 20$ {\tt
  MSTW2008} error sets, whereas for scale variations the events obtained with the scales in
(\ref{eq:centralscale}) are reweighted by the following six
combinations, $(\mu_F,\mu_R) \times \big\{(1/2,1/2), (1/2,1),
(1,1/2),(2,1), (1,2), (2,2)\big\}$, and the envelope of the
distributions constitutes the uncertainty band.

The results in figs.~\ref{fig:tev_ptt_yt}-\ref{fig:lhc_ptbj2_ybj2} are
characterised by the following pattern. In the main plot there are
three curves corresponding to \aMCatNLO+\HERWIG\ (black solid),
\POWHEG+\HERWIG\ (blue dashed) and \POWHEG+\PYTHIA\ (red dotted). In
the upper inset the relative scale (red dotted) and PDF (black dashed)
uncertainties for the \aMCatNLO+\HERWIG\ results are shown.
In the lower inset the ratio of the \aMCatNLO\ (black solid),
\POWHEG+\PYTHIA\ (red dashed), standalone \HERWIG\ (blue dotted) and
the fixed NLO (green dot-dahsed), over the \POWHEG+\HERWIG\ results
are presented.  The standalone-\HERWIG\ result has been obtained showering events
generated according to their LO cross section.

\subsection{Tevatron results}
\begin{figure}[h!]
  \begin{center}
    \epsfig{file=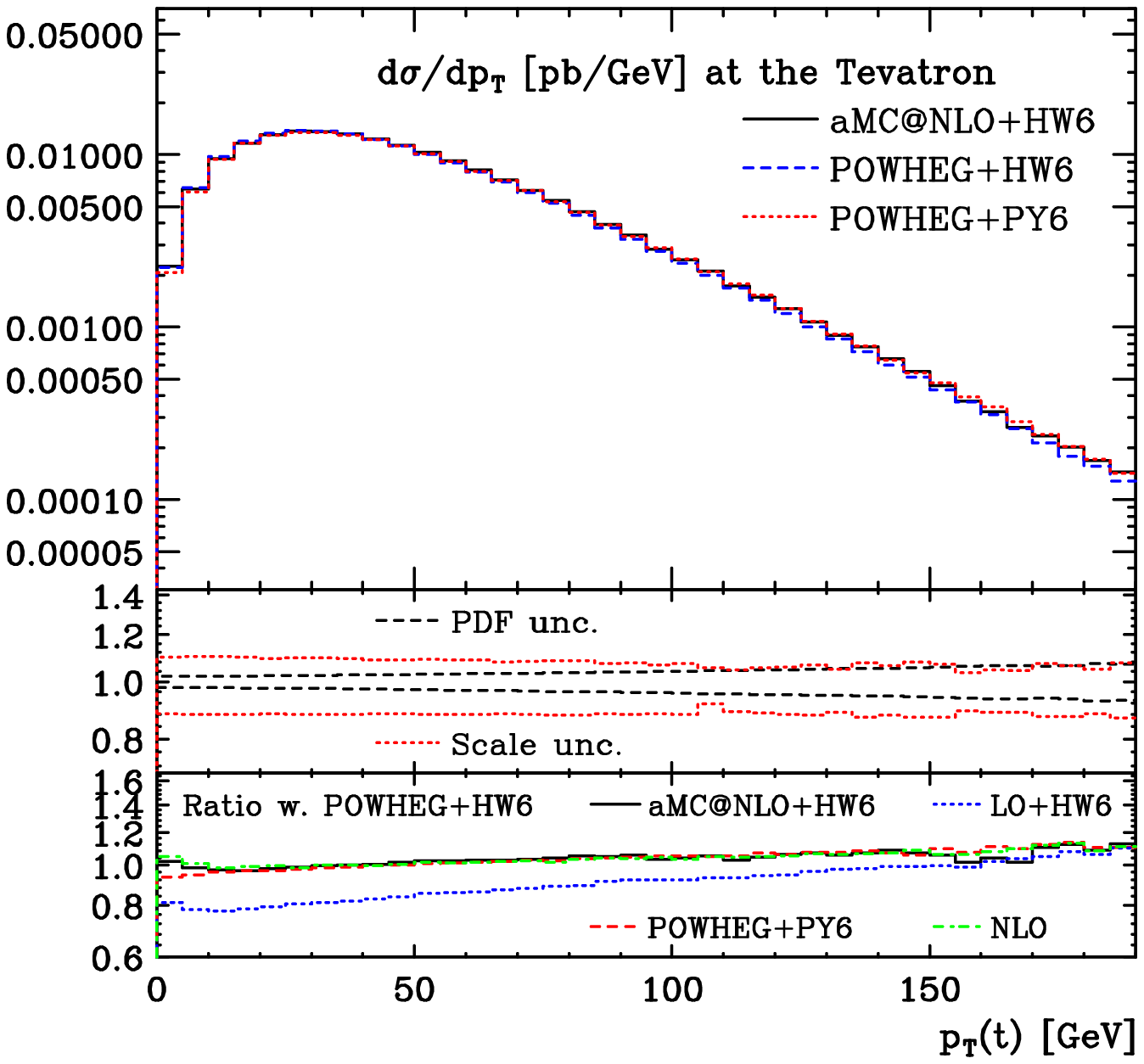,width=0.505\textwidth}~
    \epsfig{file=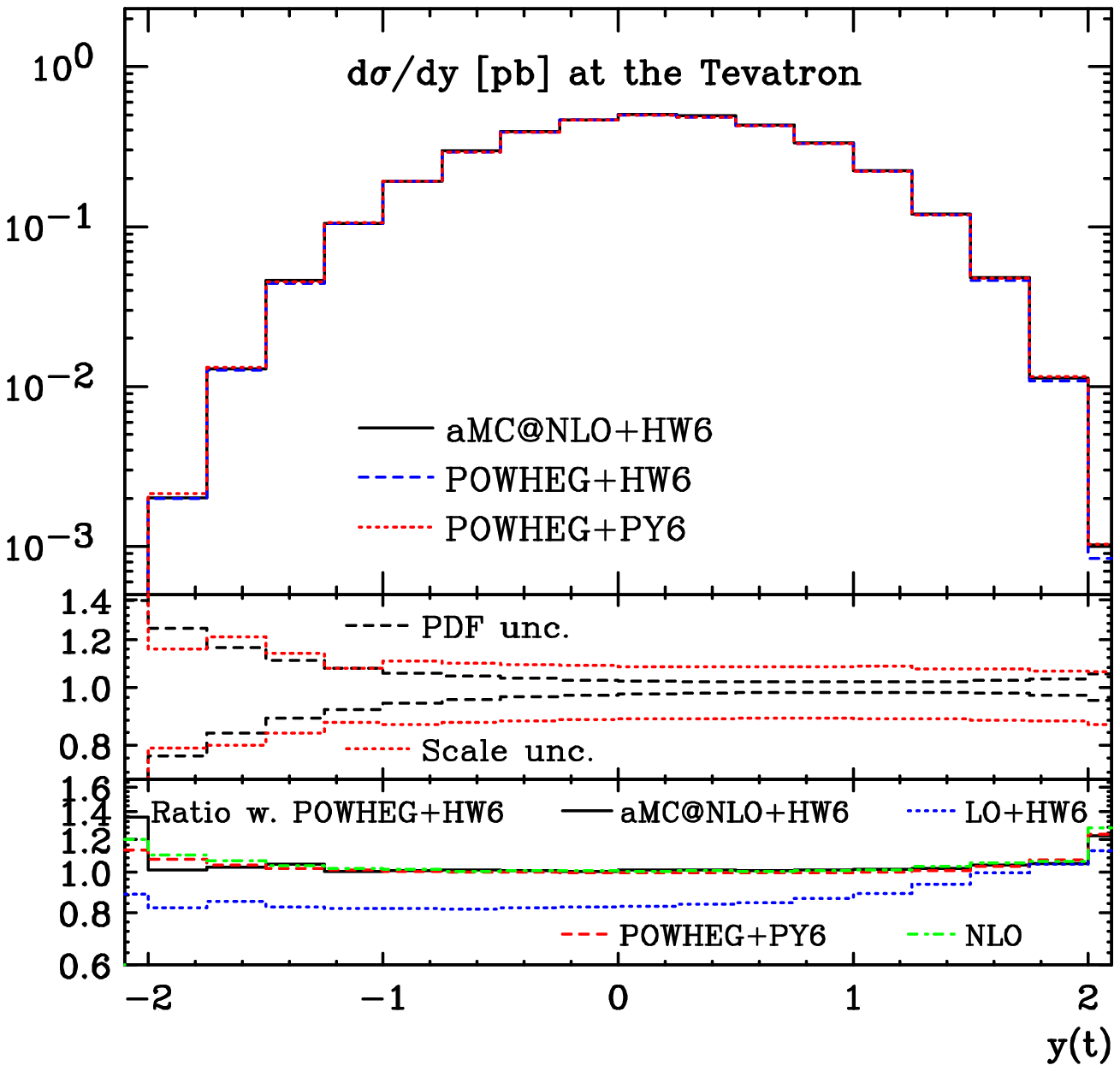,width=0.490\textwidth}
  \end{center}
  \caption{\label{fig:tev_ptt_yt} Transverse-momentum and rapidity
    distributions of the top quark at the Tevatron.}
\end{figure}

\begin{figure}[h!]
  \begin{center}
    \epsfig{file=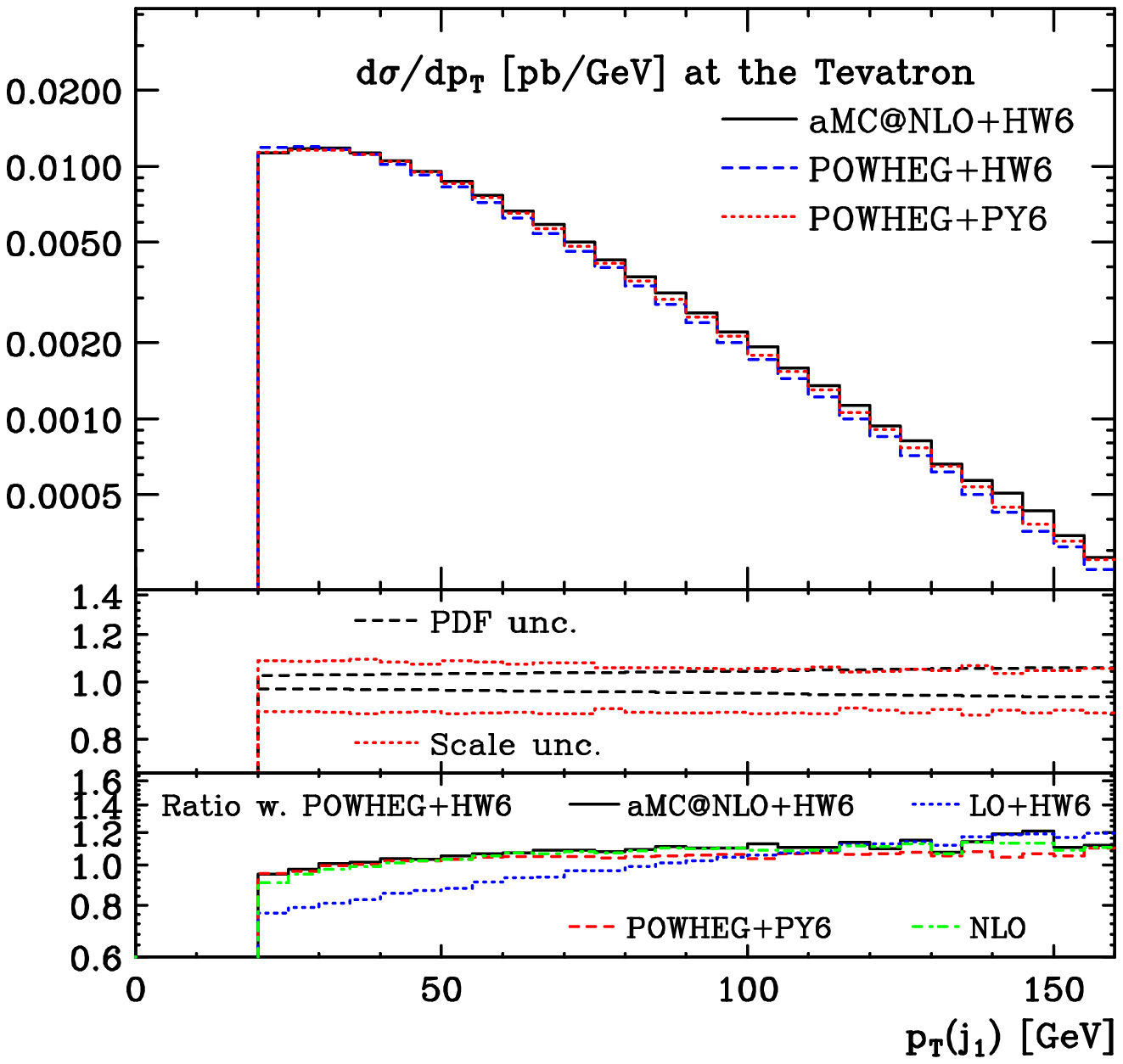,width=0.51\textwidth}~
    \epsfig{file=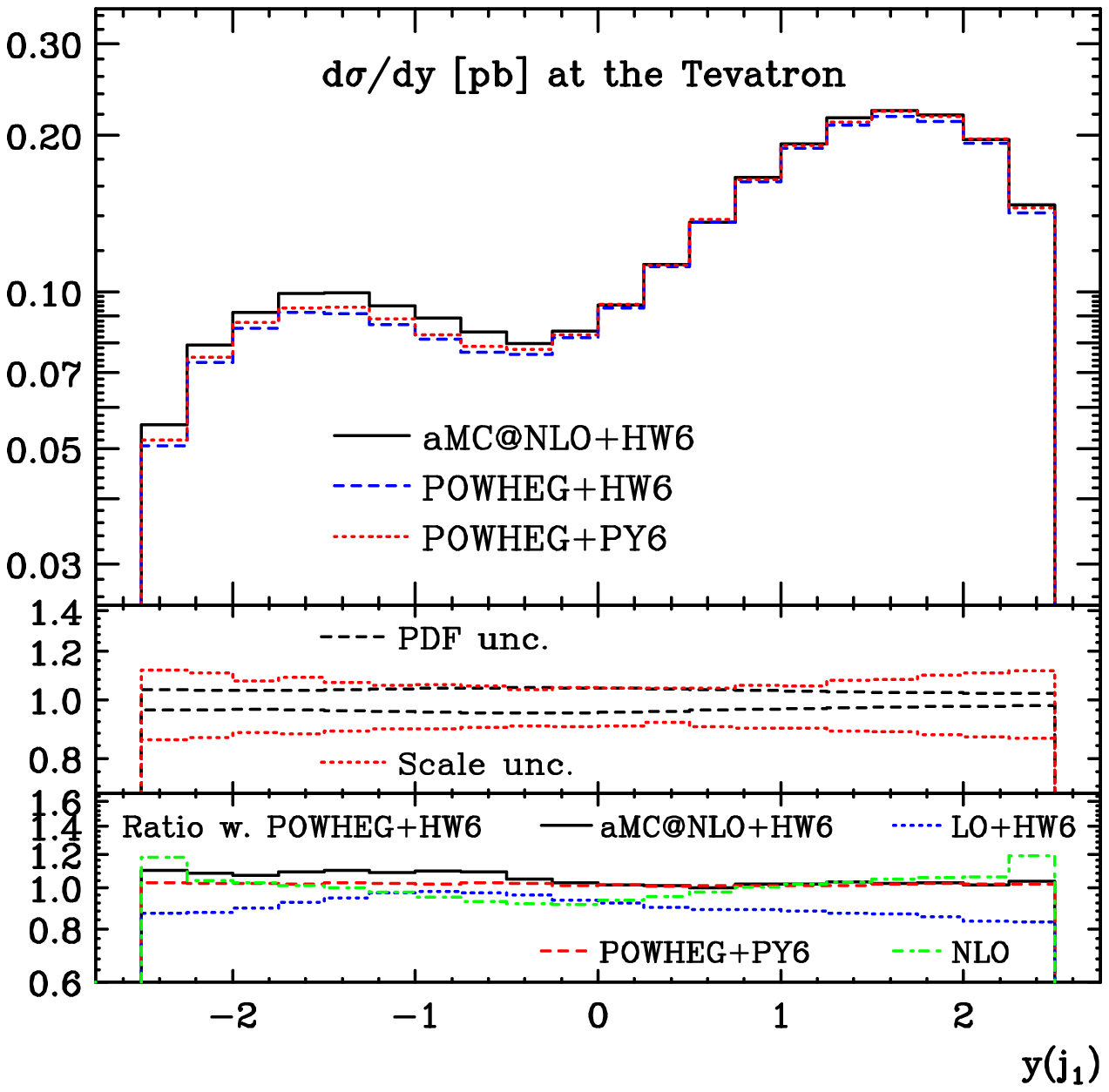,width=0.49\textwidth}
  \end{center}
  \caption{\label{fig:tev_ptj1_yj1} Transverse-momentum and rapidity
    distributions of the hardest light jet at the Tevatron.}
\end{figure}

In figs.~\ref{fig:tev_ptt_yt} and~\ref{fig:tev_ptj1_yj1} we show the
transverse momentum and the rapidity of the top quark and of the
hardest light jet, respectively. As expected, a very good agreement is
observed between the NLO and the NLO+PS results. In
  particular, all the simulations that include NLO
  corrections show an excellent agreement for observables related to the top-quark kinematics. This is slightly less the case for the hardest light jet (see, for instance, the right plot of fig.~\ref{fig:tev_ptj1_yj1}), even though the effect is still well within the uncertainty band (here shown only for the \aMCatNLO{} results): differences of the size of those appearing in fig.~\ref{fig:tev_ptj1_yj1} are however to be expected for observables affected by extra radiation. Conversely, the
discrepancy with the LO+PS curve is larger, since in the $4$-flavour scheme the NLO corrections are non-negligible, as noticed in
ref.~\cite{Campbell:2009ss}. Theoretical uncertainties, dominated by scale dependence, are in general of the order of $\pm10\%$ and constant for the observables here considered, with the exception of an increase to $\pm20\%$ for the top rapidity at large and negative values, due to the gluon-PDF uncertainty at large Bjorken $x$'s.

\begin{figure}[h!]
  \begin{center}
    \epsfig{file=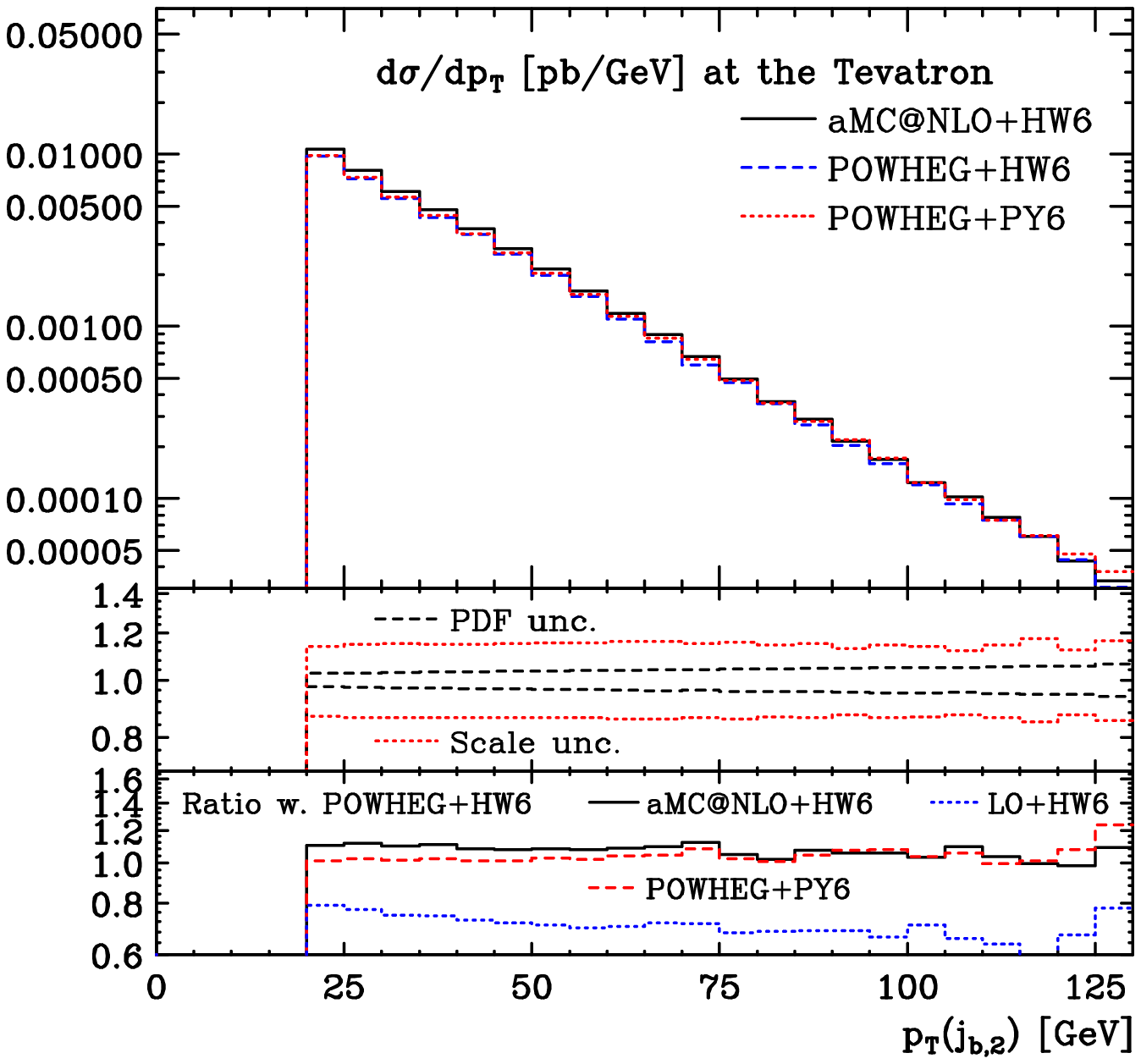,width=0.509\textwidth}~
    \epsfig{file=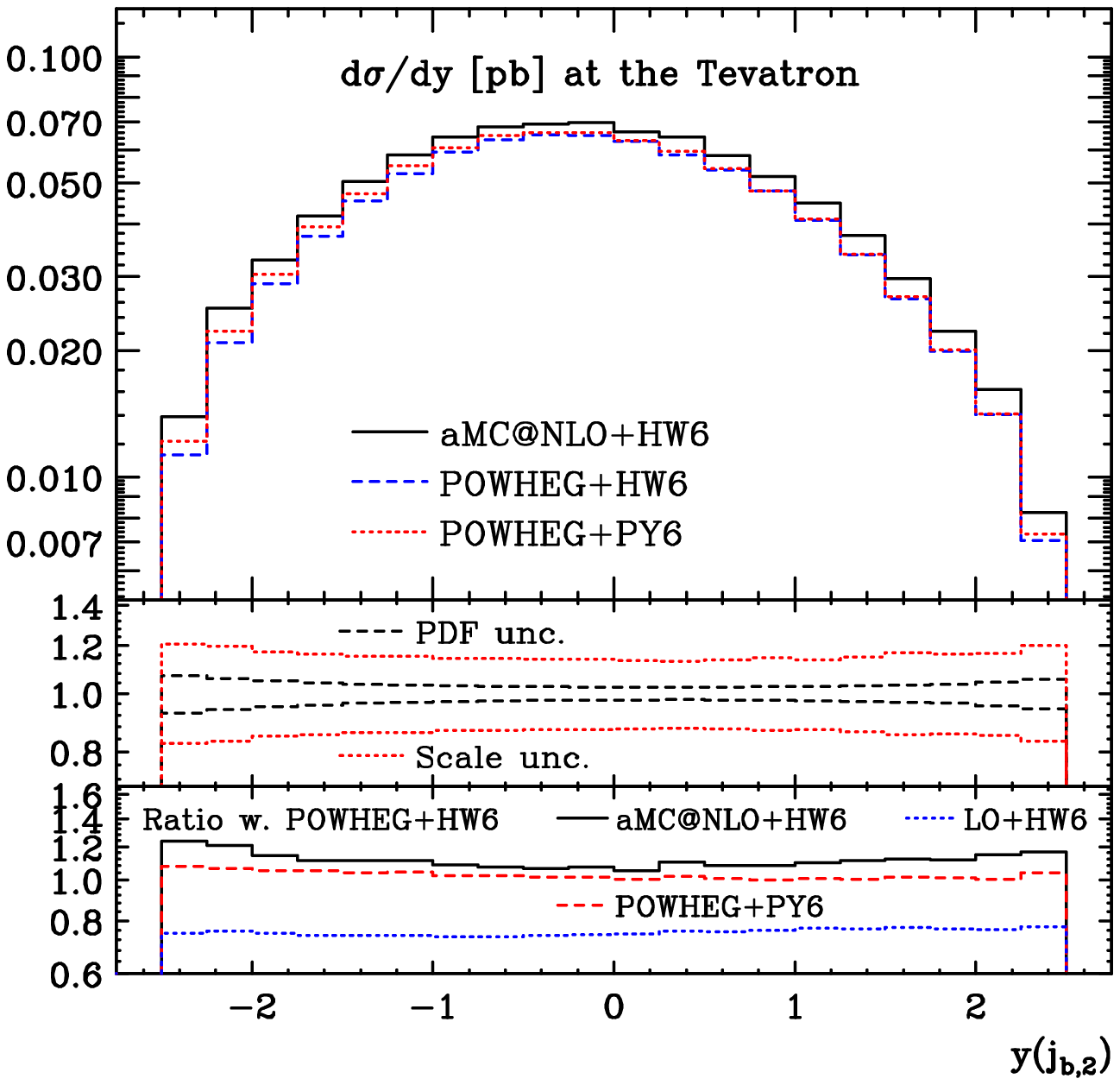,width=0.491\textwidth}
  \end{center}
  \caption{\label{fig:tev_ptbj2_ybj2} Transverse-momentum and rapidity
    distributions of the second-hardest $b$ jet at the Tevatron.}
\end{figure}

In fig.~\ref{fig:tev_ptbj2_ybj2} the transverse momentum and the
rapidity of the second-hardest $b$ jet are shown. As mentioned in the
previous sections, a good understanding of the $b$-jet kinematics is very important to discriminate between single-top and $t\bar{t}$
events. In particular, the typical single-top signature is
the presence of a hard and central $b$ jet from the top decay and of a second $b$
jet (the spectator-$b$ jet) with larger rapidity and moderate $\pt$,
arising from the initial-state $g\to b\bar{b}$ splitting, as opposite to $t\bar{t}$
events where typically two hard and central $b$ jets are present.

In our simulations, top-quark decays are not
included at the matrix-element level, therefore we have omitted the fixed-order results from these plots, as a comparison with showered predictions would not be fair: in fact, for the latter the second-hardest $b$ jet is mostly the spectator-$b$ jet,
contaminated by jets coming from the top decay, while at the parton
level there is no second-hardest $b$ jet, and the spectator-$b$ jet is always the hardest.

For NLO+PS predictions, transverse momenta are in good agreement with
each other, whereas the \aMCatNLO{} rapidity distribution is slightly
broader than the ones obtained with \POWHEG{}. However, as the
theoretical-uncertainty bands, here of the order of $\pm15\%$, are included, the \aMCatNLO{} and the \POWHEG{} results
become compatible. Residual discrepancies at this level
should be considered as an estimate of the systematic uncertainty
arising from the use of different NLO+PS-matching methods.

\subsection{LHC results}
\begin{figure}[h!]
  \begin{center}
    \epsfig{file=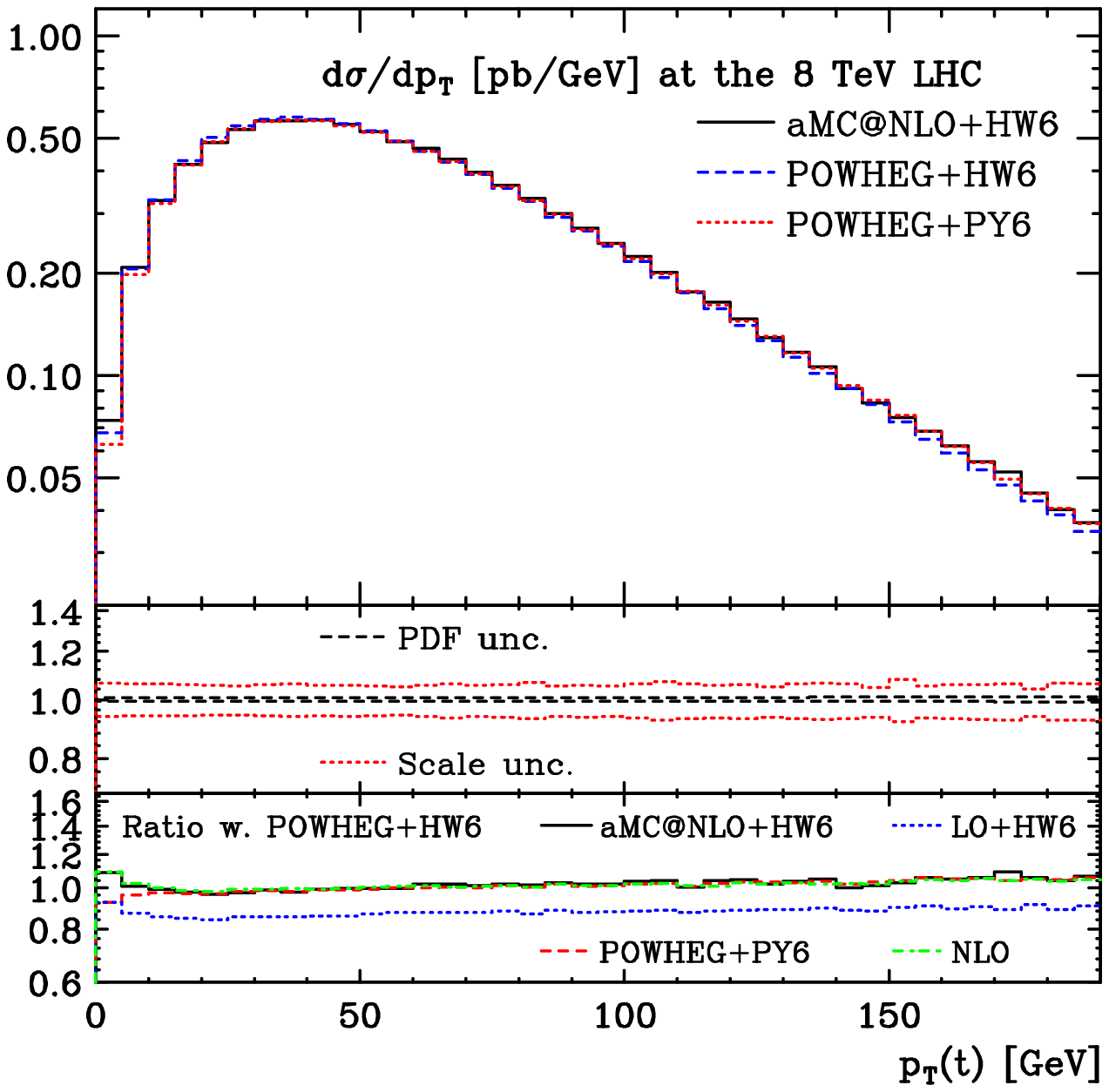,width=0.499\textwidth}~
    \epsfig{file=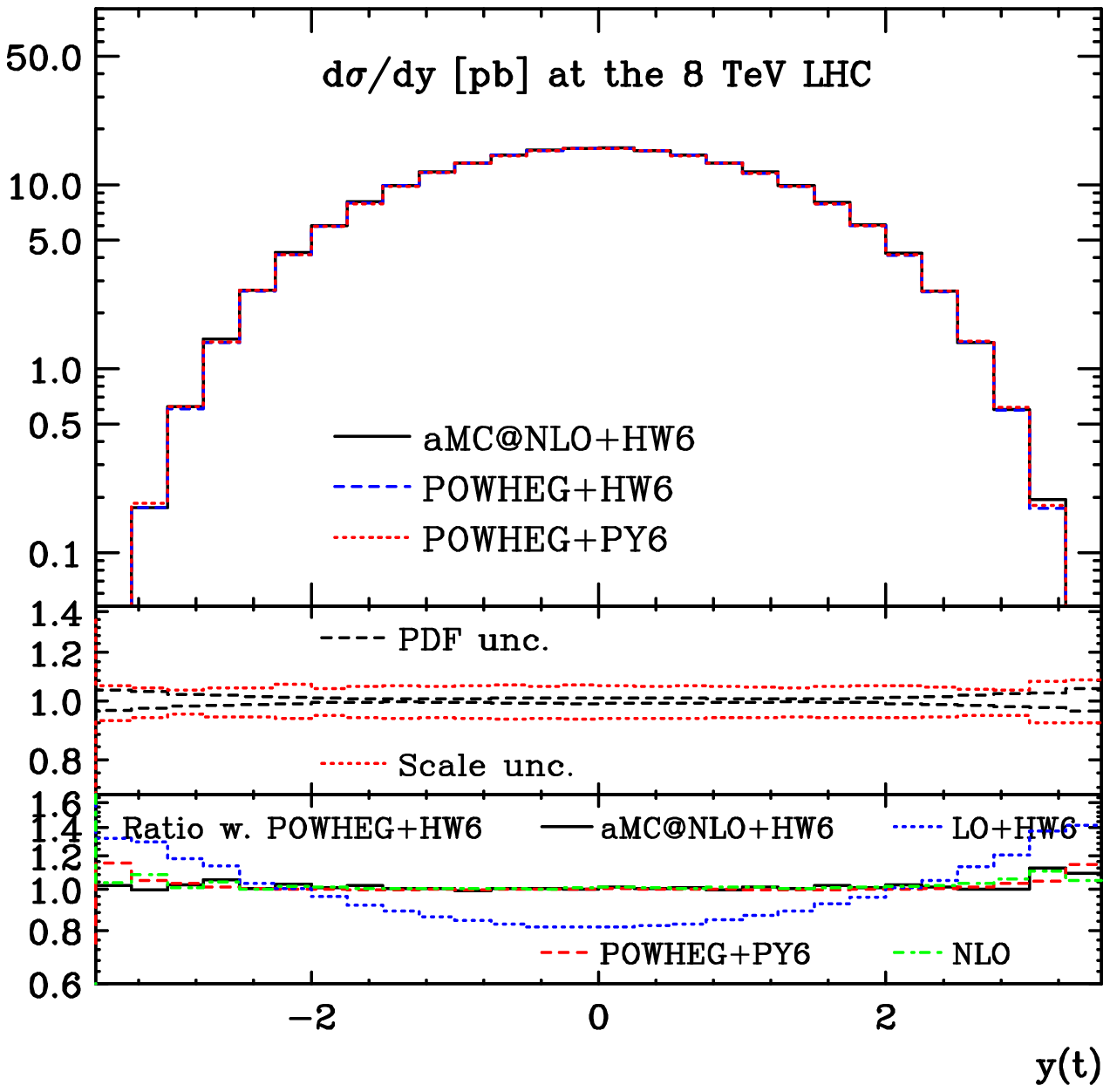,width=0.5\textwidth}
  \end{center}
  \caption{\label{fig:lhc_ptt_yt} Transverse-momentum and rapidity
    distributions of the top quark at the LHC.}
\end{figure}

\begin{figure}[h!]
  \begin{center}
    \epsfig{file=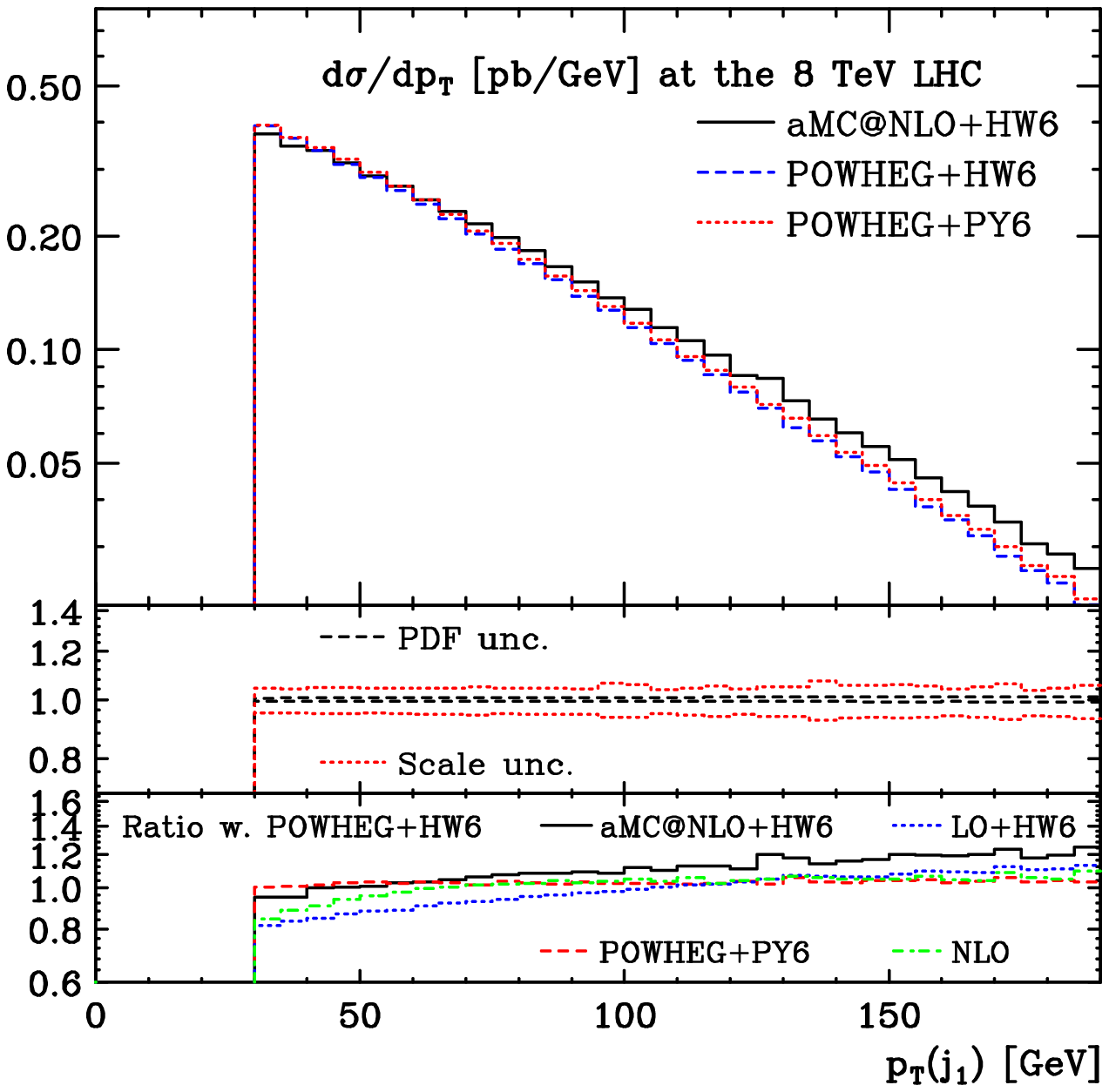,width=0.505\textwidth}~
    \epsfig{file=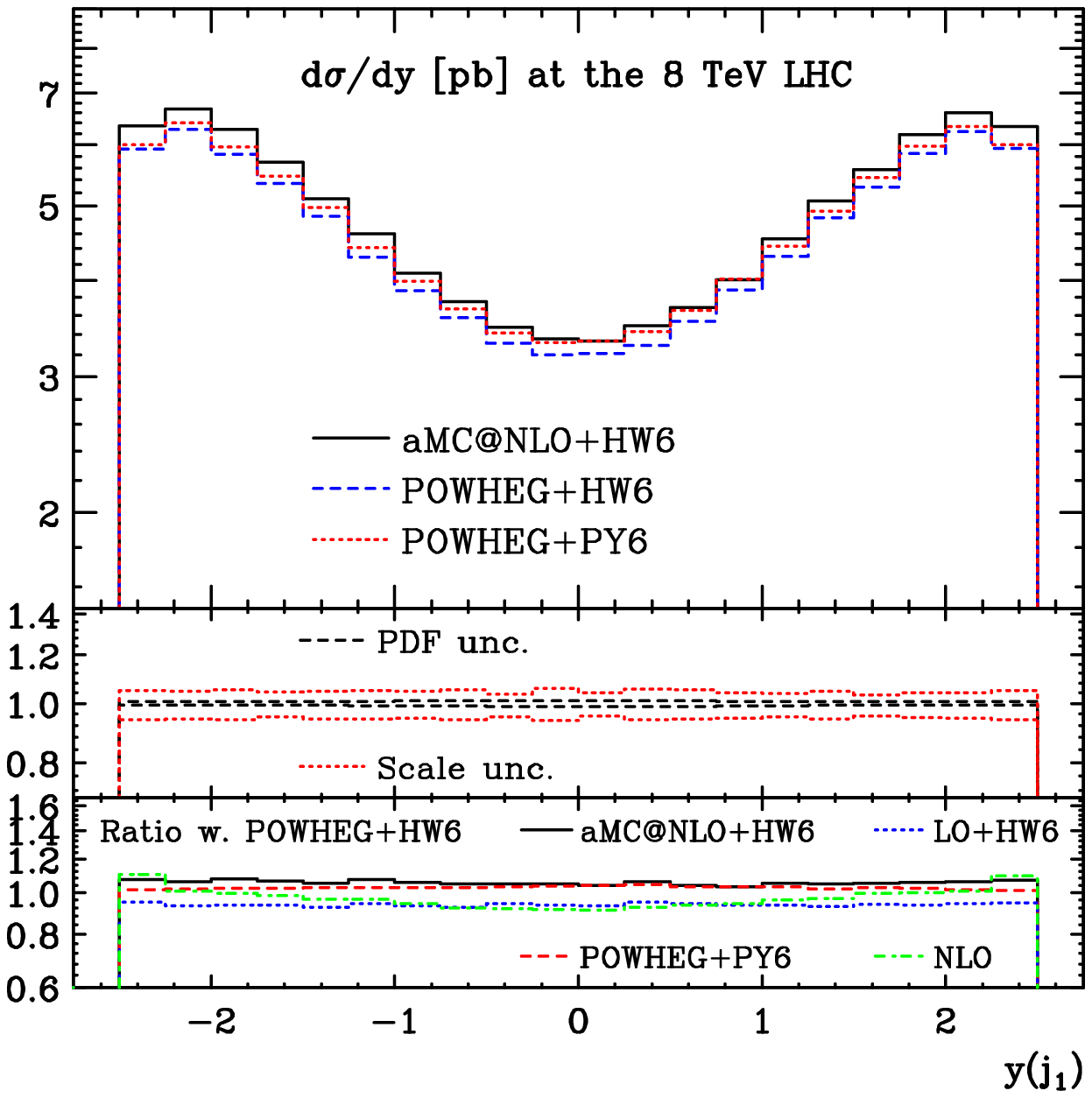,width=0.495\textwidth}
  \end{center}
  \caption{\label{fig:lhc_ptj1_yj1} Transverse-momentum and rapidity
    distributions of the hardest light jet at the LHC.}
\end{figure}

\begin{figure}[h!]
  \begin{center}
    \epsfig{file=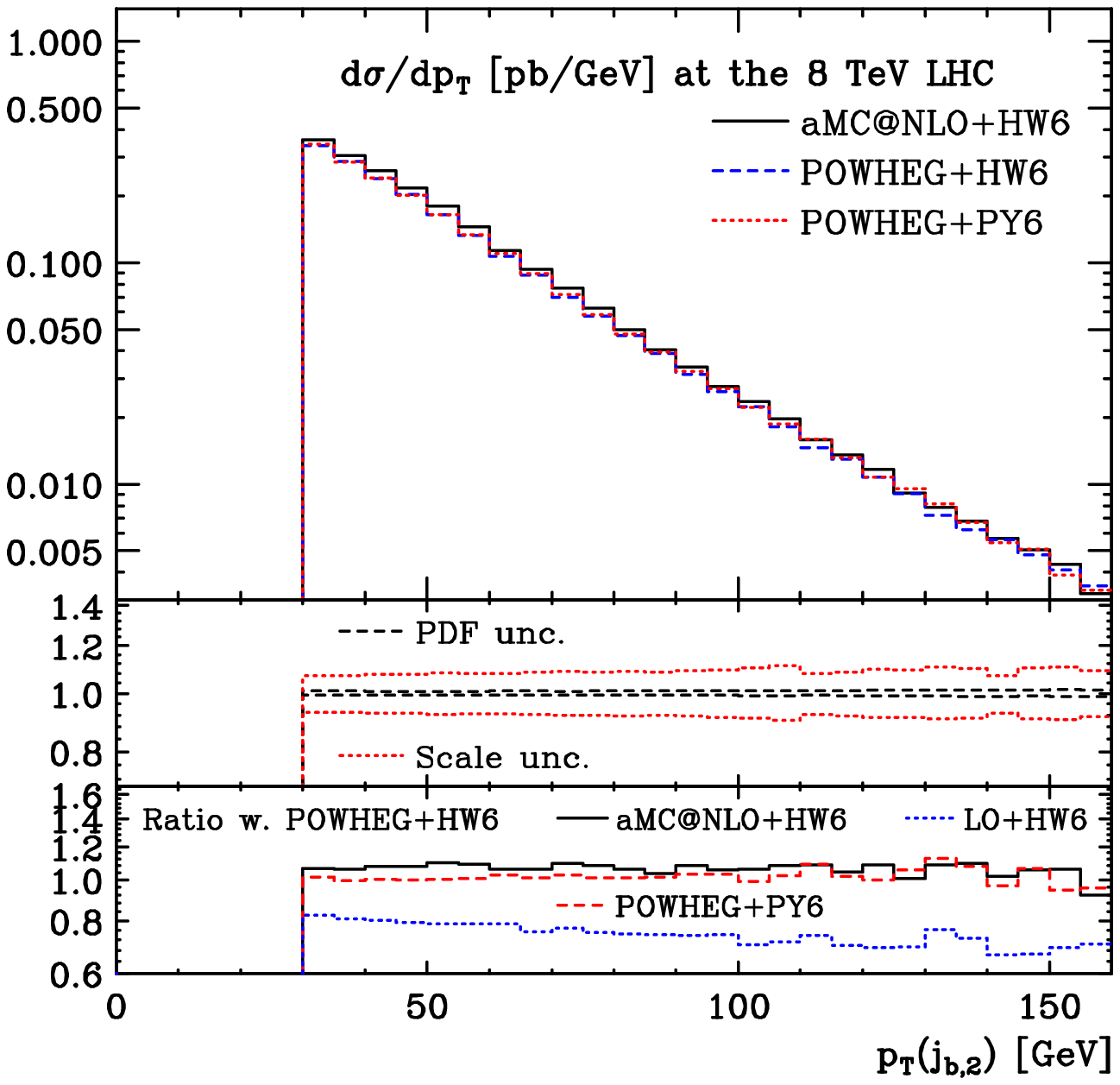,width=0.510\textwidth}~
    \epsfig{file=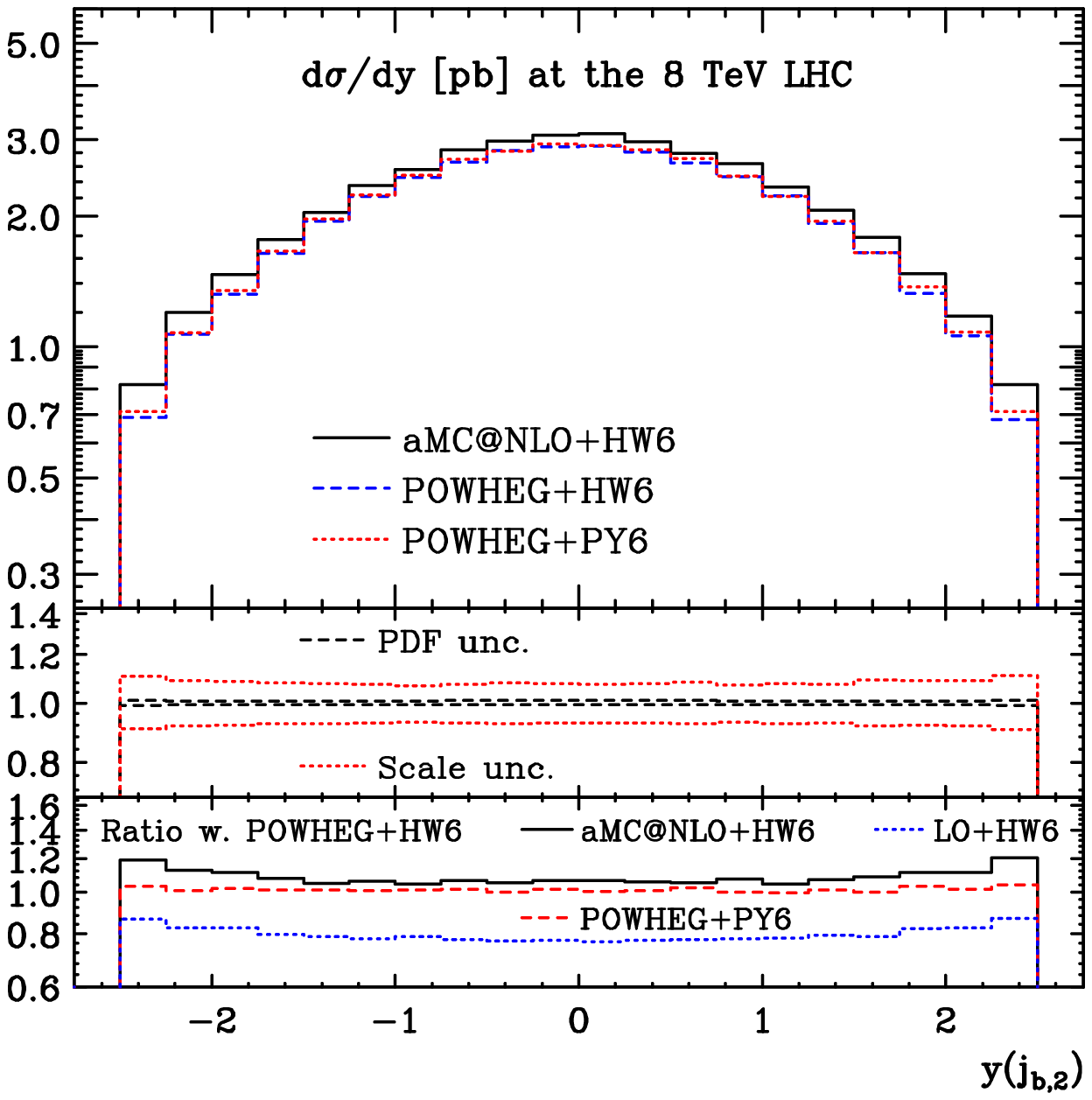,width=0.488\textwidth}
  \end{center}
  \caption{\label{fig:lhc_ptbj2_ybj2} Transverse-momentum and rapidity
    distributions of the second-hardest $b$ jet at the LHC.}
\end{figure}

In figs.~\ref{fig:lhc_ptt_yt}-\ref{fig:lhc_ptbj2_ybj2}, the same
observables as in figs.~\ref{fig:tev_ptt_yt}-\ref{fig:tev_ptbj2_ybj2} are shown for single-top production at the LHC, with a
center-of-mass energy of 8 TeV. Considerations similar to
the above ones can be drawn by inspecting these plots. One can notice
some more pronounced differences between the \aMCatNLO{} and the
\POWHEG{} results for the light jet. These are the only observables
where the disagreement between the two predictions exceeds the
theoretical-uncertainty bands. In particular, the \aMCatNLO{} result
is larger, and although the ennhancement is essentially uniform over the rapidity range examined (right plot of fig.~\ref{fig:lhc_ptj1_yj1}), a shape difference with respect to the two \POWHEG{} predictions can be seen in the $\pt$ spectrum (left plot of fig.~\ref{fig:lhc_ptj1_yj1}). However this observable, in the phase-space region considered, is sensitive to the behaviour of the parton shower in its first emission: therefore the \aMCatNLO{} result closely follows the shape of the \HERWIG{} one, while \POWHEG{} is closer to the fixed-order NLO. As commented before, in experimental analyses, these differences should be considered as a theoretical systematics of the NLO+PS-matching scheme. Scale and PDF variations are smaller than for the Tevatron, and in general constant and of the order of $\pm 5\%$ for top- and hardest-jet-related observables, and $\pm10\%$ for the spectator-$b$ jet.

As a final comment, we notice a remarkably good agreement between the
\POWHEG{} predictions obtained with the \PYTHIA{} and the \HERWIG{}
shower, for all the observables considered, both at the Tevatron and at
the LHC. It is therefore unlikely that truncated showers be relevant
in order to simulate with NLO+PS accuracy the single-top $t$-channel process in
the $4$-flavour scheme. The residual, very small dependence on the
shower model employed can be considered as a systematic effect intrinsic in the \POWHEG{} simulations.

\subsection{Comparison between the 4- and the 5-flavour MC
  simulations}
\label{sec:comparsion_4_5}

In this section we want to briefly comment on the robustness of the
NLO+PS simulation of $t$-channel single-top in the 5-flavour scheme,
and compare it with the new implementation presented in this paper.
As mentioned in the introduction, from a perturbative point of
view the 4- and the (improved) 5-flavour schemes are exactly equivalent only if all orders in perturbation theory are included, even though, as shown in refs.\cite{Campbell:2009ss,Campbell:2009gj}, the agreement is good already at the NLO level. In the following, we would like to confirm this for the NLO+PS results.

It is known~\cite{Alioli:2009je,Frixione:2010ra} that in the
simulation of single-top in the 5-flavour scheme, distributions
involving $b$-flavoured objects coming from a $g\to b\bar{b}$
initial-state splitting show unphysical pathologies at low $\pt$ and
high rapidity, when the \HERWIG{} shower is used.  The reason for this
is a simplified treatment of the non-perturbative corrections, in
particular due to a mismatch between the scale at which the backwards
shower evolution is switched off and the one at which the $b$ quarks
in the PDF are switched on.  In this respect, it is also worth
recalling that when the NLO+PS matching is performed with \MCatNLO{}
using the \HWpp{} event generator, the unphysical spikes at large
$y_{\bar{b}}$ disappear, as it was shown in~\cite{Frixione:2010ra}.
It is clear that in the $4$-flavour approach the potential
mismodelling of the $B$-hadron kinematics is fixed, because there is no
initial-state $b$ quark at the matrix-element level. Therefore we
expect the NLO+PS predictions obtained in the $4$-flavour scheme to be very solid when observables related to the spectator-$b$ jet (arising from the $B$ hadron, originated in turn by the $\bar b$ quark) are considered.
\begin{figure}[htb]
  \begin{center}
    \epsfig{file=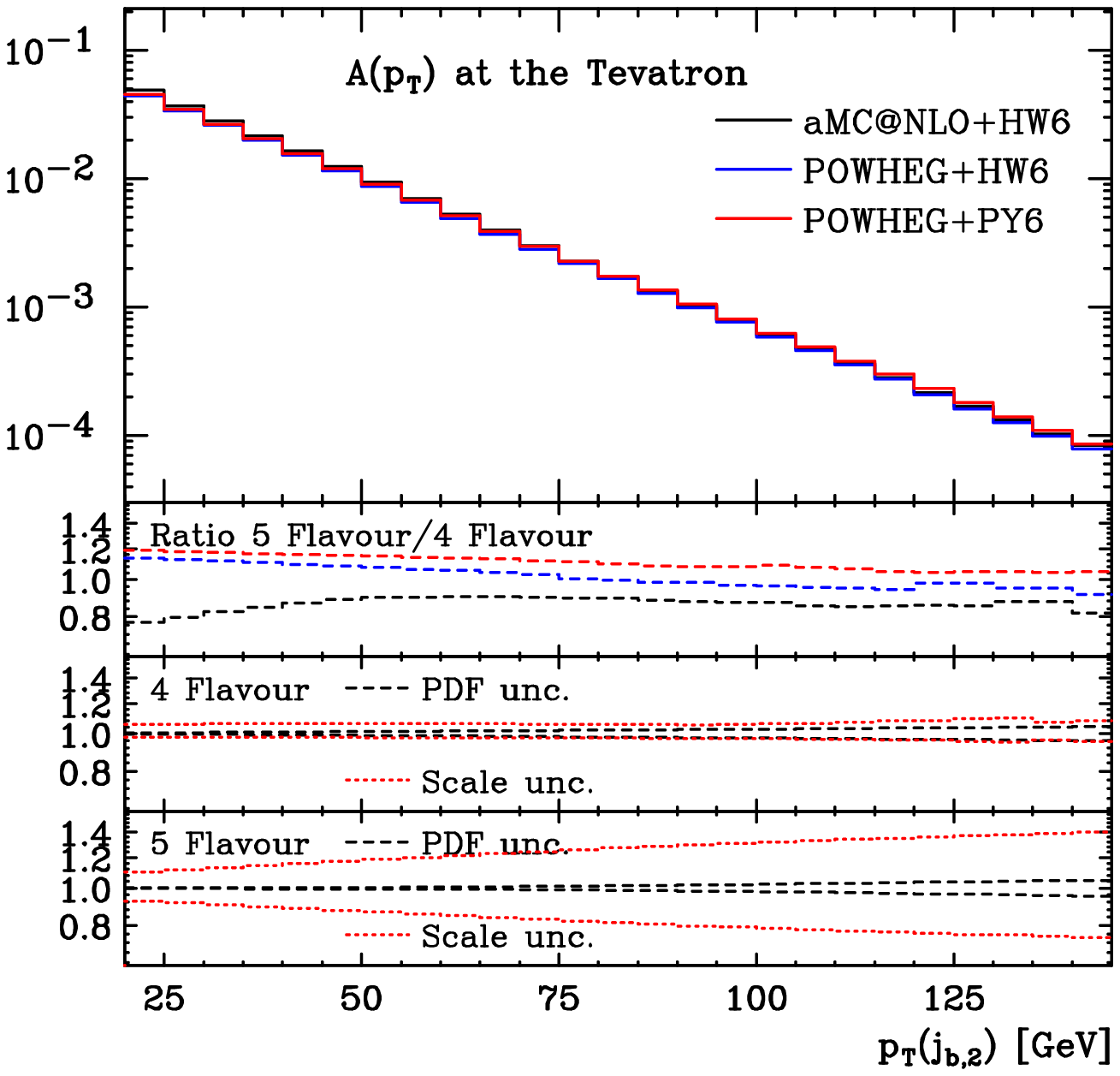,width=0.5\textwidth}~
    \epsfig{file=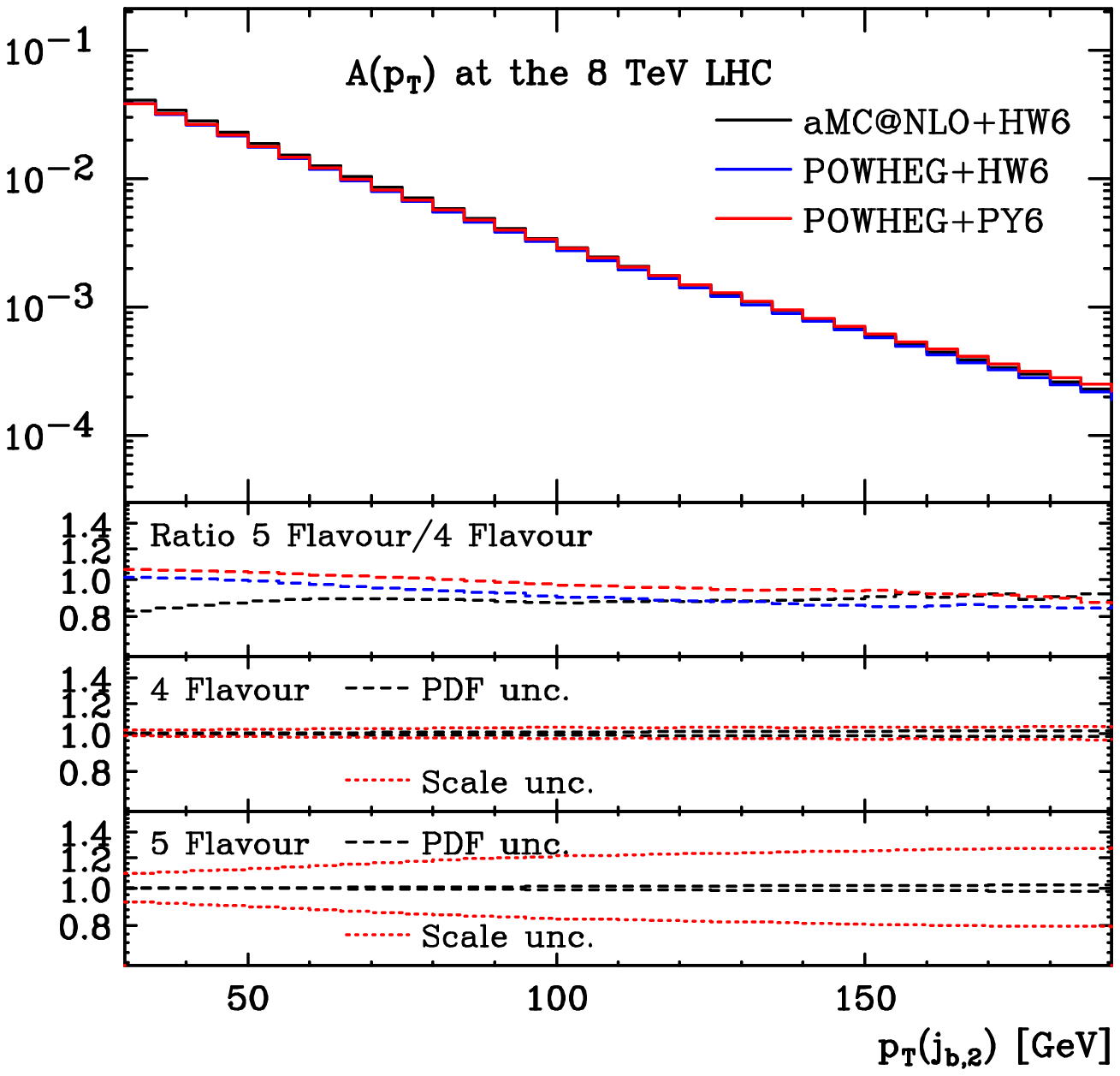,width=0.498\textwidth}
  \end{center}
  \caption{\label{fig:acceptances4v5} Acceptance as a function of the
    second-hardest $b$-jet transverse momentum, as defined in
    eq.~(\ref{eq:acceptance}).}
\end{figure}
In fig.~\ref{fig:acceptances4v5} we show the acceptance for the
spectator-$b$ jet (indentified as above with the second-hardest $b$
jet).  The acceptance is defined as
\begin{equation}
  \label{eq:acceptance}
  A(\pt)=\frac{1}{\sigma_{\sss\rm tot}} \int_{\pt}^{\infty} d{p^{(j_{b,2})}_{\sss\rm T}}\,\frac{d\sigma}{d{p^{(j_{b,2})}_{\sss\rm T}}}\,,
\end{equation}
where $\sigma_{\sss\rm tot}$ is the total NLO inclusive cross section,
and jets are required to have $|\eta_j|<2.5$. In the main panels of both plots, the NLO+PS results obtained in the $4$-flavour scheme are
shown. In the upper insets we show the ratio between the NLO+PS
acceptances computed with the $5$- and the $4$-flavour scheme, whereas
in the middle and lower insets the theoretical uncertainty for this
observable is reported, as computed with \aMCatNLO{}, using the $4$-
and the $5$-flavour scheme, respectively.\footnote{For the $5$-flavour
  predictions we have used the top mass as the central value for
  renormalisation and factorisation scales, and the {\tt MSTW2008nlo}
  PDF set.}

First of all, we notice in the main panels that the \aMCatNLO{} and
the two \POWHEG{} results are practically indistinguishable. This is encouraging, since those for observables related to the spectator-$b$ jet are genuinely NLO-accurate
predictions in the 4-flavour scheme, and therefore we expect the generators to produce results very close to each other. In the upper insets we observe 10-15\% deviations from
1 in the ratio between the $5$- and the $4$-flavour predictions. More
precisely, ratios are close to 1 at the LHC in the whole $\pt$
range, whereas differences are more pronounced at the Tevatron, and
slighlty more evident for the \aMCatNLO{} results. The smaller discrepancy at the LHC is compatible with the observations of ref.~\cite{Maltoni:2012pa}, since the Bjorken $x$'s relevant to collisions at the LHC are on average smaller than those at the Tevatron, which suppresses the logarithmic contributions that are resummed by the $b$ PDF in the 5-flavour but not in the 4-flavour total rates. However, the impact of these logarithms is quite moderate and,  when theoretical uncertainties are taken into account, the $4$- and
$5$-flavour results are compatible with each other in most of the $\pt$ range: the uncertainty of the
$5$-flavour prediction is indeed sizable (especially at the Tevatron),
as can be seen in the lower insets, since acceptances
have LO accuracy when computed in this scheme. In particular, the increase in the uncertainty at high transverse momentum is what is expected in the 5-flavour scheme for this observable, strictly related to the $\pt$ of the Born system, as discussed in \cite{Torrielli:2010aw,Frederix:2011ss}. The $4$-flavour prediction itself is instead much more accurate, as can be evinced from the middle insets.

At low transverse momentum, however, the \aMCatNLO{} ratios deviate significantly from 1 and from the \POWHEG{} results, as is evident looking at the upper insets. This behaviour is caused by the 5-flavour prediction, and it is understood as arising from the different way in which the two programs handle the $g\to b\bar b$ initial-state splitting. In particular, in the region of low $p^{(j_{b,2})}_{\sss\rm T}$, \aMCatNLO{} entirely relies on the underlying-shower description of the kinematics, thus inheriting the above-mentioned unphysical \HERWIG{} feature in the treatment of the spectator $b$. Conversely, \POWHEG{} performs the hardest emission in a way independent of the parton shower it is interfaced to, which partially prevents its predictions from this mismodelling. This interpretation is sustained by the closeness of the two \POWHEG{} curves in the upper insets.

From this comparison, one can conclude that the 4-flavour and the 5-flavour approaches are
both reliable (barred a well-known \HERWIG{} unphysical beahviour) and in mutual agreement when theoretical uncertainties
are taken into account, although, as expected, the predictions obtained
in the $4$-flavour scheme are more precise, containing NLO corrections, and more solid than the 5-flavour ones.

\section{Conclusions}
\label{sec:conclusions}
In this article we have presented for the first time two independent
computations of single-top production in the $t$ channel using the
$4$-flavour scheme, matching the QCD next-to-leading-order corrections
to parton showers.  We have used the \POWHEG{} and the \MCatNLO{}
prescriptions, as implemented in the automated frameworks \BOX{} and
\aMCatNLO{}.

We have studied observables typically considered in experimental
analyses, and compared the two NLO+PS predictions with each other as well as with fixed-order results. We have generally found very good agreement for
inclusive observables, and also for observables more sensitive to QCD
emissions, such as the light-jet and the spectator $b$-jet transverse
momenta and rapidities.  In general the predictions are compatible when theoretical uncertainties are taken into account, and the small differences encountered can be ascribed to systematic effects of the choice of different matching frameworks. We conclude therefore that the \POWHEG{} and
\MCatNLO{} results are consistent, and so are the two predictions obtained by matching \POWHEG{} with the \PYTHIA{} and the
\HERWIG{} shower.

We have also compared NLO+PS predictions obtained in the $4$- and in
the $5$-flavour schemes. It is known that NLO+PS results obtained in
the latter scheme can exhibit unphysical features in the description of
the $b$-flavoured particle produced in the fragmentation of the
$b$ quark coming from the hard process. The kinematics of this
particle needs to be predicted reliably, being relevant for the
description of the spectator-$b$ jet typically arising from it, which
in turn is important in experimental analyses. We have computed the
acceptance as a function of the spectator $b$-jet transverse momentum
in both schemes. The $4$-flavour results are the first NLO+PS-accurate predictions for this observable, and show an excellent agreement between
the \POWHEG{} and \aMCatNLO{} implementations. The differences between 4- and $5$-flavour scheme, in the region unaffected by the initial-$b$ \HERWIG{} mismodelling, are small for the acceptance, and within the theoretical uncertainty. The latter is dominated by the large error affecting the 5-flavour result, which is only LO-accurate for this observable; the $4$-flavour prediction contains instead NLO corrections, and is therefore more accurate, as we have shown explicitly.

The implementations described in this publication and/or the event
files used to produce the distributions shown here are publicly available
in the \BOX{} repository and on the \aMCatNLO{} webpage. The spin
correlations between the production and the decay of the top quark can be
included as well, which is left for future work.

\acknowledgments
E.\,R. is grateful to P.\,Nason for clarifications in the earlier stages of
this work. R.\,F. and P.\,T. thank Stefano Frixione and Fabio Maltoni for useful discussions. We also acknowledge the LHCPhenoNet network under the Grant Agreement PITN-GA-2010-264564 and the Swiss National Science Foundation under contract 200020-138206 for financial support. The work of P.\,T. is in part supported by the Swiss National Science Foundation, and in part by the ERC grant 291377,``LHCtheory: Theoretical predictions and analyses of LHC physics: advancing the precision frontier".

\bibliography{paper}
\end{document}